\documentclass[twocolumn,showpacs,preprintnumbers,amsmath,amssymb]{revtex4}
\usepackage{graphicx}% Include figure files
\usepackage{dcolumn}% Align table columns on decimal point
\usepackage{bm}% bold math
\begin{document}

\preprint{}

\title{Two-Particle Self-Consistent Approach to Anisotropic Superconductivity}% Force line breaks with \\

\author{Junya Otsuki}
\affiliation{%
$^1$Department of Physics, Tohoku University, Sendai 980-8578, Japan\\
$^2$Theoretical Physics III, Center for Electronic Correlations and Magnetism,\\
Institute of Physics, University of Augsburg, D-86135 Augsburg, Germany
}%

\date{\today}% It is always \today, today,
             %  but any date may be explicitly specified

\begin{abstract}
A nonperturbative approach to anisotropic superconductivity is developed 
based on the idea of two-particle self-consistent (TPSC) theory by Vilk and Tremblay. 
A sum-rule which the momentum-dependent pairing susceptibility satisfies is derived.
An effective pairing interaction between quasiparticles is determined 
so that the susceptibility should fulfill this exact sum-rule,
 in which fluctuations belonging to different symmetries couple at finite momentum.
% Numerical solutions reveal 
% We demonstrate 
It is demonstrated 
that the mode coupling between $d$-wave and $s$-wave pairing fluctuations leads to
suppression of the $d$-wave fluctuation near the Mott insulator.
% around half-filling.
\end{abstract}

\pacs{74.20.-z, 71.10.Fd}% PACS, the Physics and Astronomy
                             % Classification Scheme.
%\keywords{Suggested keywords}%Use showkeys class option if keyword
                              %display desired
\maketitle

\section{Introduction}
The anisotropic superconductivity originating in the Coulomb repulsion
has been a topic of continuous interest in strongly correlated electron systems.
In the typical situation described by the Hubbard model, 
the superconducting phase lies in the regime with two comparable energy scales: 
the bandwidth $W$ and the Coulomb repulsion $U$\cite{Yanase03}.
This intermediate regime is unreachable by the Monte Carlo simulations
on account of
the severe sign problem\cite{Loh90}.
Therefore, addressing the anisotropic superconductivity has been a challenging issue both in numerically and analytically.

The elemental approach is the perturbation theory with respect to $U$.
The random-phase approximation (RPA) gives an intuitive picture of the effective interaction arising from $U$\cite{Berk-Schrieffer66, Anderson-Brinkman73, Scalapino86}.
A more systematic approximation is the fluctuation exchange (FLEX) approximation, in which all physical quantities are derived from the Luttinger-Ward functional\cite{FLEX1, FLEX2}.
These theories give us basic information on the dominant pairing fluctuation in models and realistic materials\cite{Yanase03}.
In addressing the strong-correlation regime, 
$U \gtrsim W$,
however, the perturbative treatment of $U$ is not sufficient.
As a result,
the superconducting fluctuation survives close to the Mott insulator\cite{Yanase03, Takimoto02}. 
The local correlation, which is incorporated in the self energy, is responsible for the suppression of the superconducting fluctuation.

On the other hand, the strong local correlation giving rise to the Mott transition can be treated by the dynamical mean-field theory (DMFT)\cite{Metzner89, Georges}.
Its cluster extensions enable us to address the momentum-dependent vertex part which is indispensable for the anisotropic superconductivity\cite{Maier-RMP, Potthoff03}.
To exclude the size effect, 
it is necessary to take larger clusters than the minimal size\cite{Maier05}.
Other kinds of extensions without clusters
 have also been proposed to incorporate the influence of the long-range spatial fluctuations\cite{Kusunose06, Toschi07, Held08, Katanin09, Slezak09, Rubtsov09, Hafermann09}.

% There are another class of theories which start from the two-particle fluctuation in contrast to the above approaches, in which the self-energy is evaluated prior to the two-particle fluctuation.
A non-perturbative approach referred to as two-particle self-consistent theory (TPSC) has been developed by Vilk and Tremblay\cite{TPSC, Tremblay11}.
This theory starts from the two-particle fluctuation in contrast to the above approaches, 
where the self-energy is evaluated prior to the two-particle fluctuation.
In this sense, TPSC inherits the idea of the phenomenological method named self-consistent renormalization (SCR) theory, which has succeeded in describing the quantum critical phenomena\cite{Moriya-SCR, Moriya-Ueda00, Kondo-Moriya09}.
TPSC consists of two parts. 
First, an effective quasiparticle interaction is derived from the double occupancy so that the spin and charge susceptibilities satisfy the exact sum-rule. 
% The local correlation is taken into account by the double occupancy and by the Pauli principle enforced implicitly in this sum-rule. 
In this sum-rule, the local correlation is taken into account by the quantity
$\langle n_{\sigma} n_{\sigma'} \rangle$, 
which corresponds to the double occupancy for different spins and 
is reduced to $\langle n_{\sigma} \rangle$ for the same spins by the Pauli principle. 
The interaction is assumed to be independent of the energy and momentum but dependent on the spin components.
With this spin-dependence and the sum-rule, 
they have achieved a reasonable description of the spin and charge fluctuations 
taking account of the local correlation.
In the next step, the two-particle susceptibilities thus obtained are used to derive the self-energy, which therefore incorporates influence of the collective modes.

After the development,
TPSC has been applied to superconductivity.
The attractive Hubbard model has been investigated to discuss the isotropic pairing\cite{Allen01, Kyung01}.
The superconductivity arising from the repulsive interaction has been discussed by Kyung \textit{et al.}\cite{Kyung03}
The superconducting transition temperature and the single-particle properties have been discussed.
In this paper, we address the anisotropic superconductivity in a way different from ref. \cite{Kyung03}.
Namely, we extend the first step of TPSC (effective interaction) to the superconducting fluctuations.
To this end, an exact sum-rule which the pairing susceptibility satisfies is derived.
Using this sum-rule, the pairing interaction as well as the pairing susceptibility is determined.

To make the discussion concrete, we consider in this paper the two-dimensional Hubbard model
\begin{align}
% H = \sum_{ij\sigma} t_{ij} c_{i\sigma}^{\dag} c_{j\sigma}
H = \sum_{\bm{k} \sigma} (\epsilon_{\bm{k}} -\mu) c_{\bm{k} \sigma}^{\dag} c_{\bm{k} \sigma}
% + U \sum_{i} c_{i\uparrow}^{\dag} c_{i\uparrow}
%  c_{j\downarrow}^{\dag} c_{j\downarrow}
+ U \sum_{i} n_{i\uparrow} n_{i\downarrow},
\end{align}
where $\epsilon_{\bm{k}} = -2t (\cos k_x + \cos k_y) - 4t' \cos k_x \cos k_y$,
and we take $t=1$ as a unit of energy.
The system size is $N = L^2$ and 
we adopt the periodic boundary condition. 

This paper is organized as follows.
We begin with a review of TPSC in the next section.
Section~\ref{sec:TPSC-SC} describes its extension to superconductivity.
We first derive a sum-rule and then an equation for an effective pairing interaction is proposed.
Numerical results are given in Section~\ref{sec:results}.
We summarize in Section~\ref{sec:summary} with some discussions.

\section{Two-Particle Self-Consistent (TPSC) Theory}
\label{sec:TPSC}

In this section, we review TPSC by Vilk and Tremblay\cite{TPSC, Tremblay11}.
We define the susceptibility by
\begin{align}
\chi_{\sigma \sigma'}(\bm{r}, \tau)
= \langle n_{\sigma}(\bm{r}, \tau) n_{\sigma'} \rangle
 - \langle n_{\sigma} \rangle \langle n_{\sigma'} \rangle.
\end{align}
The spin and charge susceptibilities are defined from $\chi_{\sigma \sigma}$, respectively, by
\begin{align}
\chi_{\rm sp} = 2(\chi_{\uparrow \uparrow} - \chi_{\uparrow \downarrow}),
\quad
\chi_{\rm ch} = 2( \chi_{\uparrow \uparrow} + \chi_{\uparrow \downarrow}).
\end{align}
The TPSC takes account of the spin-dependence of the effective interaction.
We express this coupling constant by $U_{\sigma \sigma'}$.
We assume $U_{\uparrow \uparrow} = U_{\downarrow \downarrow}$ and 
$U_{\uparrow \downarrow} = U_{\downarrow \uparrow}$ and introduce
\begin{align}
U_{\rm sp} = U_{\uparrow \downarrow} - U_{\uparrow \uparrow},
\quad
U_{\rm ch} = U_{\uparrow \downarrow} + U_{\uparrow \uparrow}.
\end{align}
With this effective interaction, the susceptibility in the RPA is given by
\begin{align}
&\chi_{\rm sp}^{\rm (tpsc)}(q) = \frac{2\chi_0(q)}{1 - U_{\rm sp} \chi_0(q)}, \\
&\chi_{\rm ch}^{\rm (tpsc)}(q) = \frac{2\chi_0(q)}{1 + U_{\rm ch} \chi_0(q)},
\end{align}
where 
$\chi_0(q) = -(T/N) \sum_k G(k) G(k+q)$ and
$G(k) = 1/(i\epsilon_n + \mu - \epsilon_{\bm{k}})$.
We have introduced the notations
$k=(\bm{k}, i\epsilon_n)$ and
$q=(\bm{q}, i\nu_n)$ with
$\epsilon_n$ and $\nu_n$ being the fermionic and bosonic Matsubara frequencies, respectively.
The effective coupling constants are determined so that 
$\chi_{\rm sp}^{\rm (tpsc)}(q)$ and $\chi_{\rm ch}^{\rm (tpsc)}(q)$ 
satisfy the exact sum-rule. 
The self-consistency equations thus read
\begin{align}
\label{eq:tpsc1}
&\frac{T}{N} \sum_{q} \chi_{\rm sp}^{\rm (tpsc)}(q)
= n - 2 \langle n_{\uparrow} n_{\downarrow} \rangle, \\
\label{eq:tpsc2}
&\frac{T}{N} \sum_{q} \chi_{\rm ch}^{\rm (tpsc)}(q)
= n + 2 \langle n_{\uparrow} n_{\downarrow} \rangle - n^2,
\end{align}
where
$n$ is the particle number per site. 
We have assumed
$\langle n_{\uparrow} \rangle = \langle n_{\downarrow} \rangle = n/2$
and have used the relation 
$\langle n_{\sigma} \rangle = \langle n_{\sigma}^2 \rangle$,
which expresses the Pauli principle.
Provided that the double occupancy $\langle n_{\uparrow} n_{\downarrow} \rangle$ is given, 
$U_{\rm sp}$ and $U_{\rm ch}$ are determined from the above equations.
In the TPSC, the following assumption that connects 
$\langle n_{\uparrow} n_{\downarrow} \rangle$ with $U_{\rm sp}$ 
is invoked
\begin{align}
\label{eq:tpsc3}
U_{\rm sp}
= \frac{\langle n_{\uparrow} n_{\downarrow} \rangle}
{\langle n_{\uparrow} \rangle \langle n_{\downarrow} \rangle}
U.
\end{align}
This relation expresses the fact that the effective interaction is reduced in accordance with the decrease of the probability that two electrons having opposite spins exist at the same site.
Thus equations are closed, and 
$U_{\rm ch}$, $U_{\rm sp}$ and
$\langle n_{\uparrow} n_{\downarrow} \rangle$ are evaluated self-consistently for given $n$ and $U$.
We note that Eq.~(\ref{eq:tpsc3}) breaks the particle-hole symmetry and should be used only for $n \leq 1$.
The range $n>1$ is considered through the particle-hole transformation into $n<1$.

\section{Extension to Superconductivity}
\label{sec:TPSC-SC}

\subsection{Sum-rule for Pairing Susceptibility}
In the TPSC reviewed in the previous section, the sum-rule for the susceptibility plays the main role. 
To apply this idea to the anisotropic superconductivity,
we first derive a corresponding sum-rule for the superconducting fluctuations.

We consider the Cooper pair with the total momentum $\bm{q}$.
Although the present interest is on the zero-momentum pairing,
finite momentum is necessary to construct a sum-rule which the pairing susceptibility satisfies.
The annihilation operator $B_{\alpha \bm{q}}$ for the symmetry $\alpha$ is defined by
\begin{align}
B_{\alpha \bm{q}} &= \sum_{\bm{k}} g_{\alpha}(\bm{k})
 c_{\bm{k}+\frac{\bm{q}}{2} \uparrow}
 c_{-\bm{k}+\frac{\bm{q}}{2}  \downarrow} \nonumber \\
&= \sum_{ij} (g_{\alpha})_{ji} c_{i\uparrow} c_{j \downarrow}
e^{-i \frac{1}{2} \bm{q} \cdot (\bm{R}_i + \bm{R}_j)}.
\end{align}
Here
$g_{\alpha}(\bm{k})$ denotes the orbital part of the pair electrons, 
and is classified in terms of the irreducible representations of the point group.
$(g_{\alpha})_{ji}$ is the Fourier transform of $g_{\alpha}(\bm{k})$:
$(g_{\alpha})_{ji} = N^{-1} \sum_{\bm{k}} g_{\alpha}(\bm{k}) e^{i \bm{k} \cdot (\bm{R}_j - \bm{R}_i)}$.
We note that the range of the vector $\bm{q}$ is $q_x=(4\pi/L) n_x$ with $n_x=0, \cdots, L-1$, 
since the periodicity of the center-of-mass coordinate $(\bm{R}_i + \bm{R}_j)/2$ is half of $\bm{R}_i$. 
Hence, for anisotropic pairing in general, $B_{\bm{q}+\bm{G}} \neq B_{\bm{q}}$ and $B_{\bm{q}+2\bm{G}} = B_{\bm{q}}$.

The pairing dynamical susceptibility is defined by
\begin{align}
P_{\alpha \alpha'}(q)
&= \int_0^{\beta} d\tau \frac{1}{N} \langle B_{\alpha \bm{q}}(\tau) B_{\alpha' \bm{q}}^{\dag} \rangle
e^{i\nu_n \tau} \nonumber \\
&= \frac{1}{N} \sum_{\bm{k} \bm{k}'}
g_{\alpha} (\bm{k}) P(\bm{k}, \bm{k}'; q) g_{\alpha'}^*(\bm{k}'),
\label{eq:P_aa}
\end{align}
where 
\begin{align}
&P(\bm{k}, \bm{k}'; q) \nonumber \\
&= \int_0^{\beta} d\tau 
\langle 
 c_{\bm{k}+\frac{\bm{q}}{2} \uparrow}(\tau) c_{-\bm{k}+\frac{\bm{q}}{2} \downarrow}(\tau)
 c_{-\bm{k}'+\frac{\bm{q}}{2} \downarrow}^{\dag} c_{\bm{k}'+\frac{\bm{q}}{2} \uparrow}^{\dag}
\rangle
e^{i\nu_n \tau}.
\label{eq:P_kkq}
\end{align}
The off-diagonal components of $P_{\alpha \alpha'}$ are finite for $\bm{q} \neq 0$.
For $\bm{q}=0$, on the other hand,
$P_{\alpha \alpha'}$ is block-diagonal with respect to the irreducible representations in the point group.

The Fourier transform is defined by
$P(\bm{r}, \tau) = (T/N) \sum'_{\bm{q}, n} P(q) e^{i \bm{q} \cdot \bm{r} - i \nu_n \tau}$,
where the prime indicates that each element of $\bm{q}$ takes the values expressed by $q_x=(4\pi/L)n_x$ as mentioned above.
Then the equal-time local component, 
$P_{\alpha \alpha'} (\bm{r}=0, \tau=-0)$, is expressed as
\begin{align}
\label{eq:sc-sumrule}
&\frac{T}{N} {\sum_{q}}' P_{\alpha \alpha'}(q) e^{i\nu_n 0^+} \nonumber \\
&= \frac{1}{N} \sum_{ijlm}
(g_{\alpha})_{ji} (g_{\alpha'})_{ml}^*
\langle c_{m\downarrow}^{\dag} c_{l\uparrow}^{\dag} c_{i\uparrow} c_{j\downarrow} \rangle
% \hat{\delta}'_{ \frac{1}{2} (\bm{R}_i +\bm{R}_j), \frac{1}{2} (\bm{R}_l +\bm{R}_m)}
\hat{\delta}_{\bm{R}_i +\bm{R}_j, \bm{R}_l +\bm{R}_m} \nonumber \\
&\equiv {Q}_{\alpha \alpha'},
\end{align}
where $\hat{\delta}$ is Kronecker's delta extended to satisfy the periodic boundary condition.
$Q_{\alpha \alpha'}$ is block-diagonal in a manner similar to $P_{\alpha \alpha'}(\bm{q}=0, i\nu_n)$.
Eq.~(\ref{eq:sc-sumrule}) is the sum-rule which the momentum-dependent pairing susceptibility satisfies.

\begin{figure}[tb]
	\begin{center}
	\includegraphics[width=6cm]{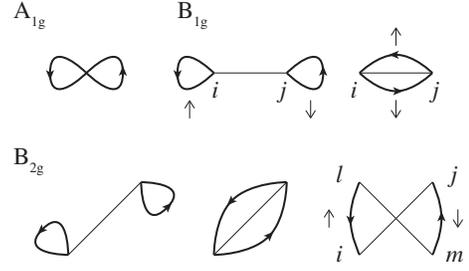}
	\end{center}
	\caption{Diagrammatic representations of ${Q}_{\alpha \alpha}$ for 
	$\alpha=$ A$_{\rm 1g}$, B$_{\rm 1g}$ and B$_{\rm 2g}$.}
	\label{fig:diagram}
\end{figure}

In the square lattice, $g_{\alpha}$ is classified by the irreducible representation of the point group D$_{\rm 4h}$. 
Table~\ref{tab:g} shows the even parity representations with the smallest $\bm{r}_{\alpha}$ in each irreducible representation. 
Here $\bm{r}_{\alpha}$ denotes one of vectors $\bm{r}_{\alpha} = \bm{R}_j - \bm{R}_i$ for which $(g_{\alpha})_{ji} \neq 0$.
\begin{table}[tb]
\begin{center}
\caption{The even-parity orbital function $g_{\alpha}(\bm{k})$ and the corresponding real-space vector $\bm{r}_{\alpha} = \bm{R}_j - \bm{R}_i$ for which $(g_{\alpha})_{ji} \neq 0$.}
\label{tab:g}
\begin{tabular}{c|c|c}
\hline
symmetry $\alpha$ & $g_{\alpha}(\bm{k})$ & $\bm{r}_{\alpha}$ \\
\hline
A$_{\rm 1g}$ & 1 & $(0,0)$ \\
A$_{\rm 2g}$ & $\sqrt{2} (\sin 2 k_x \sin 2 k_y - \sin k_x \sin 2 k_y)$ & $(2,1)$ \\
B$_{\rm 1g}$ & $\cos k_x - \cos k_y$ & $(1,0)$ \\
B$_{\rm 2g}$ & $2 \sin k_x \sin k_y$ & $(1,1)$ \\
\hline
\end{tabular}
\end{center}
\end{table}
We have chosen the normalized condition
$N^{-1} \sum_{\bm{k}} |g_{\alpha}(\bm{k}) |^2 =1$.
Explicit expressions for Eq.~(\ref{eq:sc-sumrule}) are then given by
\begin{align}
\label{eq:sc-sumrule1}
{Q}_{\rm A1g, A1g}
&= \langle n_{\uparrow} n_{\downarrow} \rangle, \\
\label{eq:sc-sumrule2}
{Q}_{\rm B1g, B1g}
&= \langle n_{\uparrow}(\hat{\bm{x}}) n_{\downarrow} \rangle 
- \langle \sigma_{+}(\hat{\bm{x}}) \sigma_{-} \rangle, \\
\label{eq:sc-sumrule3}
{Q}_{\rm B2g, B2g}
&= \langle n_{\uparrow}(\hat{\bm{x}}+\hat{\bm{y}}) n_{\downarrow} \rangle 
- \langle \sigma_{+}(\hat{\bm{x}}+\hat{\bm{y}}) \sigma_{-} \rangle \nonumber \\
&-2 \langle c_{\uparrow}^{\dag}(\hat{\bm{x}}+\hat{\bm{y}}) c_{\uparrow}(\hat{\bm{x}})
 c_{\downarrow}^{\dag} c_{\downarrow}(\hat{\bm{y}}) \rangle,
\end{align}
where $\hat{\bm{x}}$ and $\hat{\bm{y}}$ denote the primitive translation vectors.
Corresponding diagrams are shown in Fig.~\ref{fig:diagram}.
For $\bm{r}_{\alpha} \neq 0$,
$Q_{\alpha \alpha}$ consists of two-site terms and four-site terms.
When $\bm{r}_{\alpha}=(2n+1, 0)$, the four-site term is absent as in Eq.~(\ref{eq:sc-sumrule2}).
The two-site terms may be rewritten in terms of the spin and charge susceptibilities as
\begin{align}
\label{eq:two-site}
&\langle n_{\uparrow}(\bm{r}_{\alpha}) n_{\downarrow} \rangle 
- \langle \sigma_{+}(\bm{r}_{\alpha}) \sigma_{-} \rangle \nonumber \\
&= \frac{1}{4} \left[ \chi_{\rm ch}(\bm{r}_{\alpha}, \tau=0) - 3\chi_{\rm sp}(\bm{r}_{\alpha}, \tau=0) +n^2 \right]. 
\end{align}
Here we have assumed isotropy in the spin space.

The explicit forms of 
$Q_{\alpha \alpha'}$ in Eqs.~(\ref{eq:sc-sumrule1})--(\ref{eq:sc-sumrule3}) confirm
the importance of the sum-rule~(\ref{eq:sc-sumrule}).
The double occupancy $\langle n_{\uparrow} n_{\downarrow} \rangle$, appeared in Eq.~(\ref{eq:sc-sumrule1}), is one of the most important quantities in systems with the local repulsion.
For anisotropic pairings, on the other hand, 
$Q_{\alpha \alpha}$ consists of spin and charge correlations 
as the effective pairing interaction in the RPA\cite{Yanase03}.
Therefore, $Q_{\alpha \alpha}$ reflects the influence of either the local correlation or the spin correlation depending on the basis $\alpha$.

\subsection{Self-Consistent Equations}
\label{sec:equation}
We express the pairing susceptibility $P_{\alpha \alpha'}(q)$ in a RPA form phenomenologically.
To this end, we consider the following effective interaction
\begin{align}
H_{\rm int} =
\sum_{\bm{k} \bm{k}' \bm{q}} V(\bm{k} - \bm{k}')
 c_{\bm{k}+\frac{\bm{q}}{2} \uparrow}^{\dag} c_{-\bm{k}+\frac{\bm{q}}{2} \downarrow}^{\dag}
 c_{-\bm{k}'+\frac{\bm{q}}{2} \downarrow} c_{\bm{k}'+\frac{\bm{q}}{2} \uparrow}.
\end{align}
Here we assume that the coupling constant depends only on the momentum transfer $\bm{k}-\bm{k}'$.
Then, $V$ can be written in terms of the irreducible representation $g_{\alpha}(\bm{k})$ as follows:
\begin{align}
\label{eq:V_alpha}
V(\bm{k} - \bm{k}') = \sum_{\alpha} V_{\alpha}
g_{\alpha}^* (\bm{k}) g_{\alpha} (\bm{k}').
\end{align}
Here we have assumed a single basis for each irreducible representation.
Actually, $V$ may have off-diagonal elements between $g_{\alpha}$'s belonging to the same irreducible representation.

In the RPA, the two-particle Green function $P(\bm{k}, \bm{k}'; q)$ in Eq.~(\ref{eq:P_kkq}) is given by
\begin{align}
P^{\rm (eff)} & (\bm{k}, \bm{k}'; q)
= {P_0}(\bm{k}; q) \delta_{\bm{k} \bm{k}'} \nonumber \\
&- \frac{1}{N} \sum_{\bm{k}''} {P_0}(\bm{k}; q)
 V(\bm{k} - \bm{k}'') P(\bm{k}'', \bm{k}'; q),
\label{eq:P-RPA}
\end{align}
where 
\begin{align}
{P_0}(\bm{k}; q)
= T\sum_{m} G(\bm{k}+\frac{\bm{q}}{2}, i\epsilon_m) G(-\bm{k}+\frac{\bm{q}}{2}, -i\epsilon_m+i\nu_n).
\end{align}
Inserting Eqs.~(\ref{eq:V_alpha}) and (\ref{eq:P-RPA}) into Eq.~(\ref{eq:P_aa}), 
we obtain the expression for $P^{\rm (eff)}_{\alpha \alpha'}(q)$ in the RPA.
To make the notation simple, 
we use a matrix form with respect to $\alpha$ indices and denote the matrix by a hat.
Then $\hat{P}^{\rm (eff)}(q)$ is expressed as
\begin{align}
\hat{P}^{\rm (eff)} (q)
= [ \hat{P}_0(q)^{-1} + \hat{V} ]^{-1},
\end{align}
where $(\hat{V})_{\alpha \alpha'} = \delta_{\alpha \alpha'} V_{\alpha}$ and
$P_{0, \alpha \alpha'} (q)$ is defined from ${P_0} (\bm{k}; q)$ in a manner similar to Eq.~(\ref{eq:P_aa}).
We substitute $\hat{P}^{\rm (eff)}(q)$ 
into the exact sum-rule, Eq.~(\ref{eq:sc-sumrule}).
The equation reads
\begin{align}
\label{eq:tpsc-sc}
\frac{T}{N} {\sum_{q}}' \hat{P}^{\rm (eff)}(q) e^{i\nu_n 0^+}
= \hat{Q}.
\end{align}
This equation determines the effective pairing interactions $V_{\alpha}$ from 
${Q}_{\alpha \alpha'}$, which consists of double occupancy 
$\langle n_{\uparrow} n_{\downarrow} \rangle$
or equal-time correlations such as
$\chi_{\rm sp}(\bm{r}_{\alpha}, \tau=0)$
depending on the symmetry $\alpha$.

To see the tendency of the solution, let us consider the simplest situation where the off-diagonal components of $\hat{P}_0(q)$ are neglected.
Under this approximation, if ${Q}_{\alpha \alpha}$ is larger than the non-interacting value, 
the effective interaction $V_{\alpha}$ may be attractive to enhance $P^{\rm (eff)}_{\alpha \alpha}(q)$ in Eq.~(\ref{eq:tpsc-sc}).
For A$_{\rm 1g}$ symmetry, for example, the repulsive $U$ reduces ${Q}_{\rm A1g, A1g}$ in Eq.~(\ref{eq:sc-sumrule1}), leading to a repulsive effective interaction.
For B$_{\rm 1g}$ symmetry, on the other hand, 
the antiferromagnetic spin fluctuation of $\bm{q}=(\pi, \pi)$ around the half-filling
 gives a negative value for $\chi_{\rm sp}(\hat{\bm{x}}, \tau=0)$ 
to increase ${Q}_{\rm B1g, B1g}$ in Eq.~(\ref{eq:sc-sumrule2}). 
Hence the pairs with the B$_{\rm 1g}$ symmetry is enhanced as expected.

The issue of interest is thus how the solution is affected by the off-diagonal components of $\hat{P}_0(q)$, which has finite values even between different symmetries at $\bm{q} \neq 0$.
The off-diagonal susceptibility couples the pairing fluctuations in different symmetries 
and therefore,
the A$_{\rm 1g}$ and B$_{\rm 1g}$ fluctuations are not independent any more.
Numerical solutions are given in the next section with central attention on this point.

\subsection{Approximation for Equal-Time Correlations}

In the present framework, the key quantity is ${Q}_{\alpha \alpha'}$, which consists of equal-time short-range correlations.
Provided that ${Q}_{\alpha \alpha'}$ is given, 
the effective pairing interactions $V_{\alpha}$ are determined from Eq.~(\ref{eq:tpsc-sc}).
To evaluate ${Q}_{\alpha \alpha'}$, 
we may use external numerical methods such as the exact diagonalization and the quantum Monte Carlo.
Instead, we here take advantage of TPSC results reviewed in Section~\ref{sec:TPSC} so as to make the equation closed.

As shown in Eqs.~(\ref{eq:sc-sumrule2}) and (\ref{eq:sc-sumrule3}), 
${Q}_{\alpha \alpha}$ for the anisotropic pairing consists of two-site correlations and four-site correlations.
For the two-site correlations, we use the TPSC results, $\chi_{\rm sp}^{\rm (tpsc)}$ and $\chi_{\rm ch}^{\rm (tpsc)}$, in Section~\ref{sec:TPSC}.
Concerning the four-site term, we replace them by their non-interacting values.
Eliminating the four-site terms 
and using Eq.~(\ref{eq:two-site}), 
we obtain the following expression for Eqs.~(\ref{eq:sc-sumrule2}) and (\ref{eq:sc-sumrule3}):
\begin{align}
\label{eq:Q-tpsc}
{Q}_{\alpha \alpha}^{\text{(2-site)}}
&= \frac{1}{4} \left[ \chi_{\rm ch}^{\rm (tpsc)} (\bm{r}_{\alpha}, \tau=0)
 - 3\chi_{\rm sp}^{\rm (tpsc)} (\bm{r}_{\alpha}, \tau=0) \right] \nonumber \\
&+ \chi_0(\bm{r}_{\alpha}, \tau=0)
+ \frac{T}{N} {\sum_{q}}' P_{0, \alpha \alpha}(q) e^{i\nu_n 0^+}.
\end{align}

The procedure for solving Eq.~(\ref{eq:tpsc-sc}) is summarized as follows.
We first solve TPSC equations, Eqs.~(\ref{eq:tpsc1})--(\ref{eq:tpsc3}), to obtain 
$\langle n_{\uparrow} n_{\downarrow} \rangle$, $U_{\rm sp}$ and $U_{\rm ch}$.
In the next step, these results and Eq.~(\ref{eq:Q-tpsc}) are used 
as the input to Eq.~(\ref{eq:tpsc-sc}),
thus obtaining $\hat{V}$ and $\hat{P}^{\rm (eff)}(q)$ for arbitrary $q=(\bm{q}, i\nu_n)$.

\subsection{Mermin-Wagner Theorem}
We conclude this section by demonstrating that the present approach satisfies the Mermin-Wagner theorem\cite{Mermin-Wagner, Coleman73} similarly to the TPSC.
The proof in ref.~\cite{TPSC}, which concerns the magnetism, is also applicable to superconductivity.
To see this, we observe that $Q_{\alpha \alpha}$ in the right-hand side of Eq.~(\ref{eq:tpsc-sc}) is a quantity of the order of unity.
On the other hand, the left-hand side of Eq.~(\ref{eq:tpsc-sc}) diverges toward the second-order phase transition in one- and two-dimensions.
This follows from the assumption that the static component of $P^{\rm (eff)}(q)$ takes the form
\begin{align}
P^{\rm (eff)}(\bm{q} + \bm{q}_{\rm c}, 0) \sim \frac{1}{\bm{q}^2 + \xi^{-2}},
\end{align}
near the critical point,
where $\xi$ denotes the correlation length.
Therefore the superconducting phase transition of the second order is forbidden in one- and two-dimensions provided the sum-rule~(\ref{eq:sc-sumrule}) is satisfied.
This property will be confirmed numerically in the next section.

\section{Numerical Results}
\label{sec:results}

\subsection{Equal-Time Correlations}
In this section, we show numerical results for superconducting susceptibilities.
We use, as the input to Eq.~(\ref{eq:tpsc-sc}), TPSC results in Section~\ref{sec:TPSC}.
First, we check the accuracy of this input by comparing with the exact diagonalization (ED) method.

\begin{figure}[tb]
	\begin{center}
	\includegraphics[width=\linewidth]{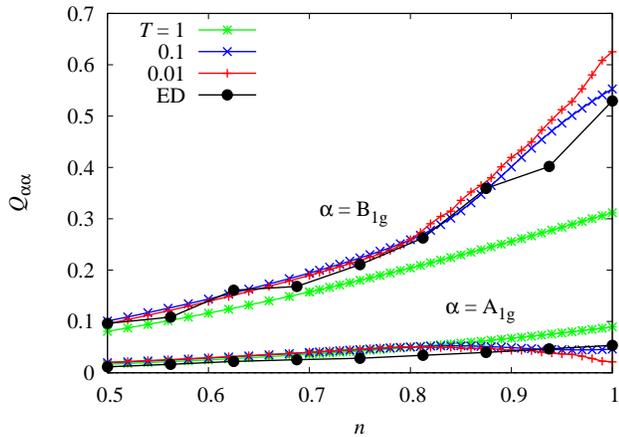}
	\end{center}
	\caption{(Color online) ${Q}_{\alpha \alpha}$ evaluated in TPSC (Eq.~(\ref{eq:Q-tpsc})) as a function of $n$ for $t'=0$, $U=8$ and $L=128$. The dots are results computed by ED with $L=4$.}
	\label{fig:chi_t0}
\end{figure}
Fig.~\ref{fig:chi_t0} shows ${Q}_{\alpha \alpha}$ for $\alpha={\rm A_{1g}}$ and B$_{\rm 1g}$ as a function of the particle number $n$ per site.
The system size $L$ is $L=128$ for TPSC and $L=4$ for ED.
The temperature dependences of TPSC results differ in the symmetries: correlations for A$_{\rm 1g}$ symmetry is suppressed with decreasing temperature, while that for B$_{\rm 1g}$ symmetry is enhanced.
These tendencies are consistent with the situation caused by the repulsive $U$ as discussed in Section~\ref{sec:equation}.
Among three values of $T$ shown in Fig.~\ref{fig:chi_t0}, $T=0.1$ is the closest to the ED results.
This correspondence is reasonable 
in view of the fact that, in the system with $L=4$, 
the energy levels of the free electrons have gaps of the magnitude 0.25.

\subsection{Static Susceptibilities}
We proceed to the results for the effective pairing interactions and the pairing susceptibilities.
The solution of Eq.~(\ref{eq:tpsc-sc}) depends on the choice of $g_{\alpha}$.
We apply the following two approximations:
\begin{itemize}
\item[(i)] {\em A$_{1g}$ or B$_{1g}$}: We neglect the off-diagonal component of $\hat{P}_0$.
In this case, the matrix equation (\ref{eq:tpsc-sc}) becomes diagonal.
\item[(ii)] {\em A$_{1g}$ and B$_{1g}$}:
We solve the 2 by 2 matrix equation with $\alpha={\rm A_{1g}}$ and B$_{\rm 1g}$.
\end{itemize}
The approximation (ii) is the minimal choice to include both the local correlation and the spin fluctuation.
Comparison of results in the above two approximations makes it clear the influence of the off-diagonal component $P_{0, \alpha \alpha'}$,
which couples pairing fluctuations in different symmetries, 
in the present equation.
We shall not present results including other symmetries and long-range pairings,
since they make only a quantitative change.

Fig.~\ref{fig:Veff} shows the effective pairing interactions $V_{\alpha}$ as a function of $n$.
The result in the approximation (i) agrees with the analysis in Section~\ref{sec:equation}:
$V_{\rm A1g}$ is repulsive because of suppression of 
$\langle n_{\uparrow} n_{\downarrow} \rangle$, 
while $V_{\rm B1g}$ is attractive due to the enhancement of the antiferromagnetic correlation between spins on the nearest neighbor sites.
The attraction in this approximation, however, turns out to be inaccurately much enhanced so that the superconducting fluctuation predominates the antiferromagnetic fluctuation 
(this situation becomes clearer in Fig.~\ref{fig:phase_diagram}).
Inclusion of the off-diagonal component of $\hat{P}_0$ suppresses the attraction in B$_{\rm 1g}$ symmetry.
We can see, in approximation (ii), that
$V_{\rm B1g}$ turns to repulsive, the extent of which is conspicuous around $n=1$.
Nevertheless, $V_{\rm B1g}$ still has tendency to attraction, 
making a minimum around $n=0.9$ as indicated by the arrow in Fig.~{\ref{fig:Veff}.
\begin{figure}[tb]
	\begin{center}
	\includegraphics[width=\linewidth]{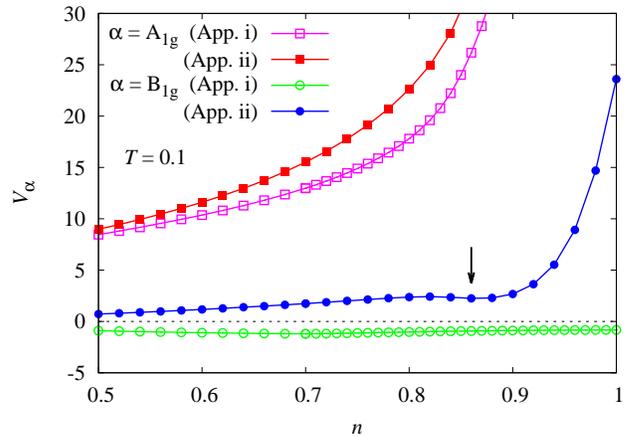}
	\end{center}
	\caption{(Color online) The effective pairing interaction $V_{\alpha}$ at $T=0.1$ for $t'=0$, $U=8$ and $L=128$. The arrow indicates a minimum of $V_{\rm B1g}$ in approximation (ii).}
	\label{fig:Veff}
\end{figure}

With the effective interaction $V_{\alpha}$ shown above, 
$P_{\alpha \alpha'}^{\rm (eff)}(q)$ for arbitrary $\bm{q}$ and $\nu_n$ can be evaluated.
We here show results only for $\bm{q}=\bm{0}$ and $\nu_n=0$ in B$_{\rm 1g}$ symmetry and discuss the tendency to the superconductivity.
The inverse of $P_{\alpha \alpha}^{\rm (eff)}(\bm{0}, 0)$ with $\alpha={\rm B_{1g}}$
is shown in Fig.~\ref{fig:chi_B1g}.
\begin{figure}[tb]
	\begin{center}
	\includegraphics[width=\linewidth]{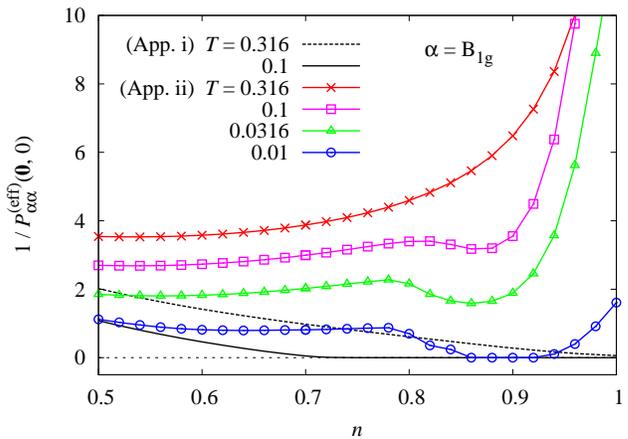}
	\end{center}
	\caption{(Color online) The inverse of the static susceptibility $P^{\rm (eff)}_{\alpha \alpha}(\bm{0}, 0)$ with $\alpha={\rm B_{1g}}$ as a function $n$ for $t'=0$, $U=8$ and $L=128$.}
	\label{fig:chi_B1g}
\end{figure}
In the approximation (i), the susceptibility is nearly diverged in a wide range of $n\gtrsim 0.7$ at $T=0.1$.
In actual, it does not diverges exactly, since the present theory satisfies the Mermin-Wagner theorem.
Compared to this result, the susceptibility in the approximation (ii) is strongly suppressed, and its inverse first touches the zero at $T\simeq 0.01$.
It is remarkable that, in contrast to the approximation (i),
 the B$_{\rm 1g}$ fluctuation does not go toward divergence around $n=1$,
accompanying the strong suppression of the A$_{\rm 1g}$ fluctuation.

\begin{figure}[tb]
	\begin{center}
	\includegraphics[width=\linewidth]{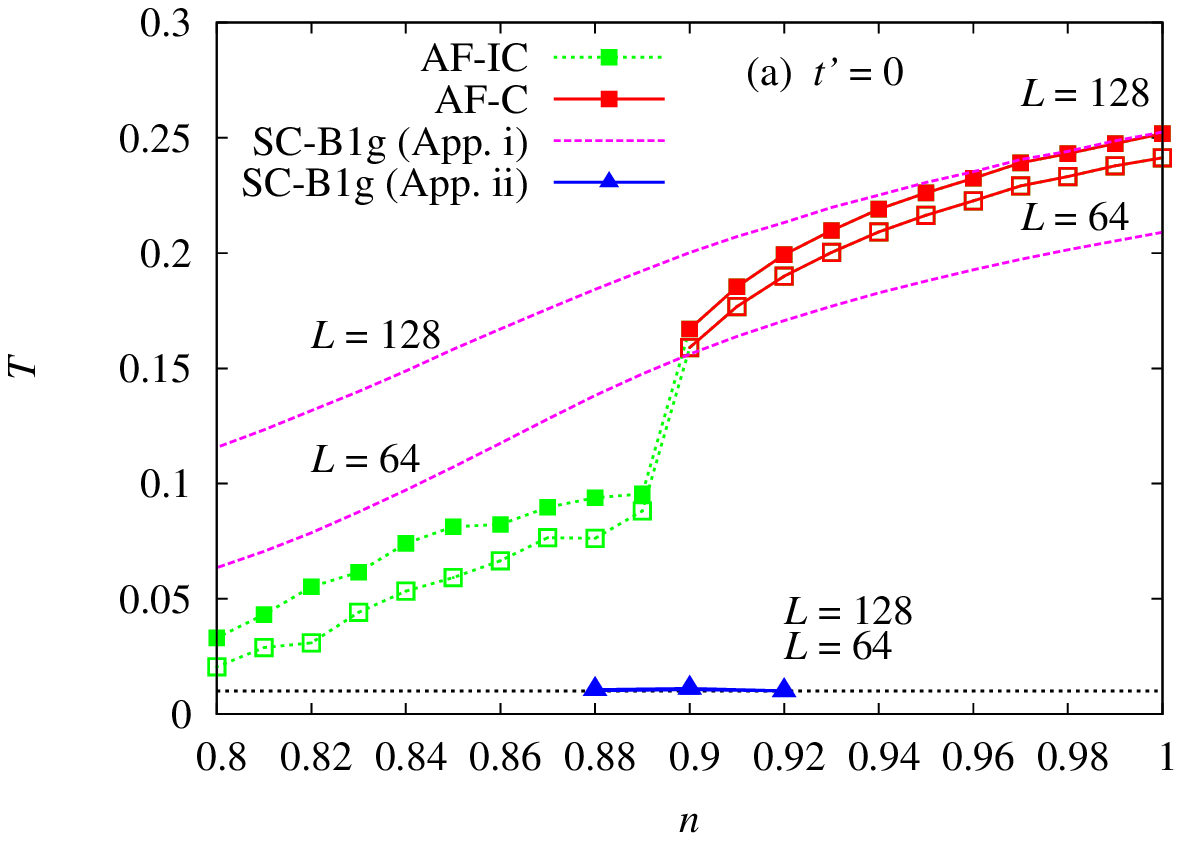}
	\includegraphics[width=\linewidth]{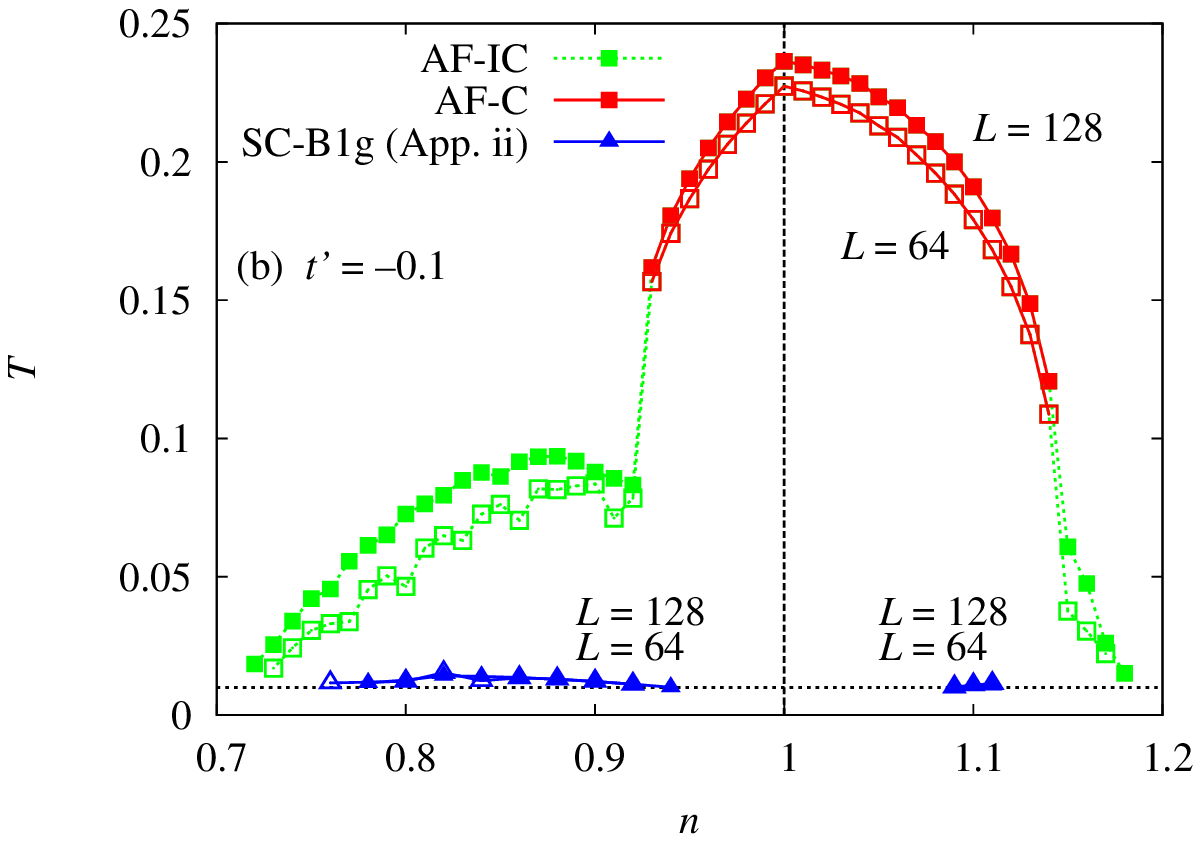}
	\end{center}
	\caption{(Color online) A phase diagram of the ``quasi-two dimensional" system calculated in the present theory (B$_{\rm 1g}$ superconductivity) and in TPSC (magnetism) for $U=8$. AF-C indicates the antiferromagnetism with $\bm{q}=(\pi, \pi)$ and AF-IC indicates ordering with the vector other than $(\pi, \pi)$ The horizontal dotted line indicates the lowest temperature $T=0.01$ in the calculation. The system size is $L=128$ (closed symbol) and $L=64$ (open symbol). (a) $t'=0$ and (b) $t'=-0.1$.}
	\label{fig:phase_diagram}
\end{figure}
Although there is no phase transition in the present calculation in two dimensions, 
to see the tendency of the fluctuations, 
we define a ``transition temperature" as the temperature at which the susceptibility equals $10^3$.
We refer to the resultant phase diagram as ``quasi-two dimensional" phase diagram hereafter.
The phase diagram for $t'=0$ is shown in Fig. \ref{fig:phase_diagram}(a). 
The antiferromagnetic transition temperature $T_{\rm AF}$ calculated in TPSC is also plotted.
The superconducting transition temperature $T_{\rm c}$ in the approximation (i) is comparable to $T_{\rm AF}$.
In the approximation (ii), on the other hand, the superconducting phase appears only around $n=0.9$ as is expected from $V_{\rm B1g}$ in Fig.~\ref{fig:chi_B1g}.
However, it turns out that $T_{\rm c}$ is always smaller than $T_{\rm AF}$.
This apparently unreasonable result is ascribed to the separate evaluation of $T_{\rm AF}$ and $T_{\rm c}$, which will be discussed in the next section. 
We also show the ``quasi-two dimensional phase diagram'' for $t'=-0.1$, 
which is expected to be a typical situation in the cuprate superconductors\cite{Yanase03}.
The range $n>1$ has been evaluated by transforming into $n<1$.
The cusp at $n=1$ is due to the use of Eq.~(\ref{eq:tpsc3}), which breaks the particle-hole symmetry.
We can see an enlargement of the superconducting phase in hole-doped regime while it is comparable to or even narrower than that of $t'=0$ in electron-doped regime.
However, the problem that $T_{\rm c}$ does not overcome $T_{\rm AF}$ still remains in this parameter.

\section{Summary and Discussions}
\label{sec:summary}

We have extended TPSC to anisotropic superconductivity.
The sum-rule, Eq.~(\ref{eq:sc-sumrule}), is the key equation,
 which relates the superconducting susceptibility to the 
spin and charge correlations expressed by $Q_{\alpha \alpha'}$.
The effective pairing interactions are determined so that the RPA-type phenomenological susceptibility satisfies the exact sum-rule.
We remark that,
in the present matrix equation, 
the off-diagonal susceptibility $P_{0, \alpha \alpha'}(q)$ plays an important role
in coupling the superconducting fluctuations in different symmetries at finite $\bm{q}$.
As a result of this ``mode coupling", 
the $d$-wave pairing fluctuation is suppressed around half-filling accompanying the $s$-wave pairing fluctuation, which is strongly suppressed by the local repulsion $U$.

Describing the suppression of the $d$-wave fluctuation near half-filling has been one of issues 
in the perturbation theories.
An essential factor lacking there is the local correlation effect, 
which the self-energy correction is one of factors responsible for.
In the present equations, on the other hand,
the suppression is achieved by the coupling with the $s$-wave fluctuation, which directly reflects the local repulsion $U$. 
In other word, the local correlation effect is incorporated within two-particle quantities through the mode coupling between superconducting fluctuations in different symmetries. 

Although the $d$-wave fluctuation is indeed suppressed around half-filling, $T_{\rm c}$ seems inaccurately small compared to $T_{\rm AF}$. 
This difference is ascribed to the structure of the present equations.
In the present framework, $\chi_{\rm sp}$ and $\chi_{\rm ch}$ are first calculated to obtain $T_{\rm AF}$ 
and with use of them, the pairing fluctuations are evaluated. 
Hence, the feedback from the superconducting fluctuation to the spin fluctuation is not taken into account.
This consideration brings us a possible improvement of the present theory: 
to treat the spin and superconducting fluctuation equally, an anisotropic vertex part giving rise to the unconventional spin-density wave\cite{Ikeda98} should be included,
and we may further bring the parquet formalism to treat both the fluctuations on an equal footing\cite{Dominicis-Martin64, Nozieres-Parquet, Kusunose10}

Finally, we again remark the importance of the sum-rule Eq.~(\ref{eq:sc-sumrule}) for pairing susceptibility.
Theories which satisfies this sum-rule follows the Mermin-Wagner theorem.
The perturbation theories such as RPA and FLEX as well as DMFT and its cluster-extensions do not satisfy this sum-rule.
This sum-rule could be one of directions in developing a framework addressing the anisotropic superconductivities.

\begin{acknowledgments}
The author would like to thank H. Kusunose for useful discussions and comments on the manuscript.
He is also indebted to
J. Schmalian, T. Takimoto, D. Vollhardt and H. Yokoyama for stimulating discussions
and J. Nasu for a technical advice.
The author is supported by JSPS Postdoctoral Fellowships for Research Abroad.
\end{acknowledgments}

\end{document}